\documentclass[runningheads]{llncs}
\usepackage[T1]{fontenc}
\usepackage{lscape}
\usepackage{enumitem}
\usepackage{tablefootnote}
\usepackage{array}
\usepackage{hyperref}
\usepackage{fancyhdr}

\usepackage{graphicx}
\begin{document}
\newcolumntype{P}[1]{>{\centering\arraybackslash}p{#1}}
\title{Quality Characteristics of a Software Platform for Human-AI Teaming in Smart Manufacturing}
\titlerunning{Quality Characteristics for Human-AI Teaming}
%
\author{Philipp Haindl\inst{1} \and
Thomas Hoch\inst{1} \and
Javier Dominguez\inst{2} \and
Julen Aperribai\inst{2} \and
Nazim Kemal Ure\inst{3} \and
Mehmet Tunçel\inst{3}
}
\authorrunning{Haindl et al.}
%
\institute{Software Competence Center Hagenberg, Hagenberg, Austria 
\email{\{philipp.haindl,thomas.hoch\}@scch.at}\\ \and
IDEKO Research Center, Elgoibar, Spain\\
\email{\{jdominguez,japerribai\}@ideko.es}\\
\and
Istanbul Technical University, Istanbul, Turkey\\
\email{\{ure,tuncelm\}@itu.edu.tr}}

\maketitle              

\begin{abstract}
As AI-enabled software systems become more prevalent in smart manufacturing, their role shifts from a reactive to a proactive one that provides context-specific support to machine operators. In the context of an international research project, we develop an AI-based software platform that shall facilitate the collaboration between human operators and manufacturing machines. 

We conducted 14 structured interviews with stakeholders of the prospective software platform in order to determine the individual relevance of selected quality characteristics for human-AI teaming in smart manufacturing. These characteristics include the ISO 25010:2011 standard for software quality and AI-specific quality characteristics such as trustworthiness, explicability, and auditability. The interviewees rated trustworthiness, functional suitability, reliability, and security as the most important quality characteristics for this context, and portability, compatibility, and maintainability as the least important. Also, we observed agreement regarding the relevance of the quality characteristics among interviewees having the same role. On the other hand, the relevance of each quality characteristics varied depending on the concrete use case of the prospective software platform. 

The interviewees also were asked about the key success factors related to human-AI teaming in smart manufacturing. They identified improving the production cycle, increasing operator efficiency, reducing scrap, and reducing ergonomic risks as key success criteria. In this paper, we also discuss metrics for measuring the fulfillment of these quality characteristics, which we intend to operationalize and monitor during operation of the prospective software platform.

\end{abstract}
\section{Introduction}
The applications of AI in smart manufacturing are numerous, ranging from improving maintenance times for machinery to detecting defects in the machine or the product to preventing injury to workers. In general, collaborative processes in smart manufacturing are characterized by alternating phases of reactive and proactive elements, with each actor supporting the other alternately. AI-enabled smart manufacturing systems are capable of self-sensing, self-adapting, self-organizing, and self-decision \cite{qu_smart_2019,phuyal_challenges_2020}, enabling them to respond to physical changes in the production environment in a variety of ways. AI-guided interactions in the manufacturing process might include stopping machines, adapting production tasks, or suggesting a change in production parameters. Achieving effective teaming between machine operators and AI-enabled manufacturing systems, however, requires mutual trust based primarily on self-sensing and self-adaptation of each actor. 
\setcounter{footnote}{0}
In the frame of the EU-funded Teaming.AI project\footnote{https://www.teamingai-project.eu}, we develop a software platform that allows for human-AI teaming in smart manufacturing. While we already presented a reference architecture in \cite{icse2022}, in this work we elaborate on the individual relevance of different quality characteristics towards such a software platform. For this purpose we conducted 14 structured interviews with different stakeholders of the prospective platform in which they rated the individual relevance of 11 different quality characteristics. A further objective of our study was the identification of key success factors and metrics that serve to evaluate the fulfillment of these quality characteristics during development and operation of the platform.

The remainder of the paper is structured as follows: Subsequently, we sketch the research context of the project in Section \ref{sec:researchcontext}, before we elaborate on the current state of research in this field in Section \ref{sec:relatedwork}. Afterwards, in Section \ref{sec:researchquestions} we describe our research questions and the used methodology in this study. Following to that, in Section \ref{sec:results} we present the results of this study and discuss our findings in Section \ref{sec:discussion}. Subsequently, in Section \ref{sec:threats} we describe possible threats to the validity of our study and conclude our paper in Section \ref{sec:conclusion}.

\section{Research Context}
The research consortium of our project consists of six research and development centers and universities, three specialized SMEs for software development of AI-based software systems, two industry partners in the automotive industry for plastic injection of car components, and one industry partner for wind power plant assembly. One key contribution of this research project is the develoment of an AI-based software platform for human-AI teaming in smart manufacturing. In the following, we describe the use cases (UC) of the three industry partners that shall be supported by this software platform.

\label{sec:researchcontext}

\subsection{UC1: Quality Inspection}
Our first industry partner manufactures injection molded components for the automotive industry. The main objective of this use case is to support the machine operator during visual quality inspection. The software platform shall classify products as OK or not-OK (NOK) with the latter being double-checked by the machine operator. Therefore, it interacts with the machine operator during quality inspection and fault analysis and provides context-specific information for adjusting parameters in order to mitigate product defects.

\subsection{UC2: Parameter Optimization}
The second industry partner also produces plastic parts for the automotive industry and its use case focuses primarily on optimizing injection parameters. To this end, the software platform should predict possible process deviations and identify likely root failure causes. Thereby, it can provide explanations for its findings (e.g. likelihoods), and the machine operator can provide feedback to the software platform. As opposed to the previous use case, in this instance visual quality inspection is performed by the machine operator. Moreover, in this use case, the software platform shall monitor the interaction between the machine operator and the injection machine as well as analyze its sensor data and parameters in order to detect process deviations prematurely.

\subsection{UC3: Large-Scale Parts Assembly}
The third industrial partner specializes in high-precision manufacturing of large-scale parts used in wind turbines, as an example. In this time-consuming production process, automated and manual tasks are incorporated. Both of these production tasks are characterized by high variability in their execution times, making task management challenging. The software platform should identify manual tasks associated with milling operations of large-scale parts and collect information about the estimated time for each of these tasks. With its tracking system, the software platform can determine the location of the machine operator. By combining this information with context information, such as machine data, the software platform acts as a mediator between the milling machine and the operator. Therefore, it should (a) improve communication between the operator and the machine, (b) allow rescheduling of similar assembly tasks, for example, combining automatic milling tasks with manual tasks, and (c) perform an ergonomic risk assessment of two simultaneous tasks as regards static loads.

\subsection{Stakeholder Roles}
\label{roles}
In the following we describe the different stakeholder roles with their exemplary activities, which were identified during requirements engineering. 

\begin{itemize}
\item \textbf{Data Protection Officer (DPO):} Ensures that a company respects the laws protecting individuals' personal data (e.g., the GDPR, by controlling the processing of data and auditing the system.
\item \textbf{Software Scientist (SS):} Queries runtime data of the software components of the software platform, e.g., logging information, for evaluating and optimizing the behaviour of the system.
\item \textbf{Data Scientist (DS):} Applies statistical methods onto data processed by the software platform, e.g., parameter tuning of ML components.
\item \textbf{Machine Operator (MO):} Visually inspects the produced parts, clamping and adjusting the workpieces or performing manual tasks on the machine, e.g., obtaining measurements and making parameter adjustments.
\item \textbf{Production Line Manager (PLM):} Monitors and optimizes the processes for producing and assembling the product or its parts on the shopfloor.
\end{itemize} 

\section{Related Work}
\label{sec:relatedwork}
We separate the current state of research related to our study into three streams. The first stream focuses on quality requirements of ML-based software systems. Based on a qualitative interview study with ten requirements engineers, Habibullah and Horkoff \cite{habibullah_non-functional_2021} explored the engineers' experiences and perceptions of quality requirements for ML-based software systems. The study shows that most engineers in industrial settings have difficulties formulating quality requirements for ML-based software systems. This often leads to quality requirements neither being organized, prioritized, nor effectively monitored during the development of such systems. Vogelsang and Borg \cite{vogelsang_requirements_2019} interviewed four data scientists using semi-structured questionnaires to examine how they elicit and specify functional and quality requirements of ML-based software systems. The authors stress that it is vital to understand ML-related performance measures to state good functional requirements for such systems. Also, systems must be designed from the beginning in such a manner that additional requirements towards explainability, trustworthiness, or even specific legal requirements can later be implemented with moderate effort. Horkoff \cite{horkoff_non-functional_2019} examined requirements engineering (RE) practices for eliciting quality requirements towards ML-based software systems. The author states that researchers and users of ML-based software systems lack an effective methodology to express and specify quality requirements for ML-based software systems, including targets and trade-offs, e.g., based on domain-specific best practices. Khan et al. \cite{khan_handling_2022} reviewed current RE methodologies for eliciting and documenting quality requirements for ML-based IIoT systems. To this end, the authors compared SysML, GORE-MLOps \cite{ishikawa_evidence-driven_2020}, and Pinto's RE methodology \cite{pinto_requirement_2021} for autonomous systems. The paper stresses the lack of a generic RE methodology for elicitating quality requirements of ML-based software systems. 

The second stream of related works examines quality characteristics of ML-based software systems. Siebert et al. \cite{siebert_towards_2020} presented a categorization of quality characteristics complemented by an operational software quality model for ML-based software systems. The definition and relevance of the quality characteristics is based on a literature-based review, complemented by workshops with industrial partners. The quality model allows to objectively assess the adherence to quality requirements throughout the development of ML-based software systems. An important prerequisite for the operationalization of the quality characteristics relates to their decomposition using metrics which can be measured throughout the engineering cycle of such systems. Lenarduzzi et al. \cite{lenarduzzi_software_2021} elaborated a method for identification of quality issues in ML-based software systems, gathered from experience reports of their research group and self-ethnography. According to the authors, root failure causes for the most frequent quality issues can be attributed to six groups, ranging from lack of developer skills, deficiencies in development and test processes, model version incompatibilities, and communication problems. They argue that training software developers is the most efficient way to mitigate quality issues in ML-based systems.

Finally, the third stream of related works focuses on quality assurance and quality models for ML-based software systems. Fujii et al. \cite{fujii_guidelines_2020} conducted a survey to evaluate the usefulness of quality guidelines for ML-based software systems. These quality guidelines address the handling of quality characteristics, test architecture, and test viewpoints for different domains. The authors criticize that the analyzed guidelines do not address the integration of explainability tools in the engineering activities of ML-based software systems. The authors assume that in practice this often leads to disregarding the quality assurance of explainability requirements or conducting it incompletely. Kuwajima et al. \cite{kuwajima_engineering_2020} studied quality models for safety-critical ML-based software systems. Therefore they analyzed the gaps between the ISO 25010:2011 (SQuaRE) standard \cite{25010:2011} for software quality and quality characteristics relevant for ML-based software systems. Their results show that the quality requirements towards machine learning models are often vaguely specified, which in turn negatively affects their interpretability and robustness. Felderer and Ramler \cite{felderer_quality_2021} analyzed terminology and challenges for quality assurance of AI-based software systems along the perspectives of artifact type, process, and quality characteristics. In total, they identified eight key challenges for this context, e.g., understandability and interpretability of AI models, accuracy and correctness measures, or the handling of quality requirements in AI-based software systems.

\section{Research Questions and Methodology}
\label{sec:researchquestions}
Our study started with the definition of candidate scenarios \cite{sutcliffe_scenario-based_2003,sutcliffe_supporting_1998} that encompass the context and the anticipated functionality from the stakeholders' perspectives when interacting with the prospective software platform. These scenarios were originally defined by our research group and were therefore only described at a high level of abstraction. Based on these candidate scenarios we designed an interview-based case study to (a) refine these scenarios into more fine-grained functional requirements, (b) assess the completeness of the scenarios to fully cover the required functionality of the software platform, (c) assess each of 11 quality characteristics in terms of its importance to the overall platform from the stakeholders' perspective, and (d) elicit the key success criteria related to the software platform.
\begin{figure}[h!]
\includegraphics[width=\textwidth]{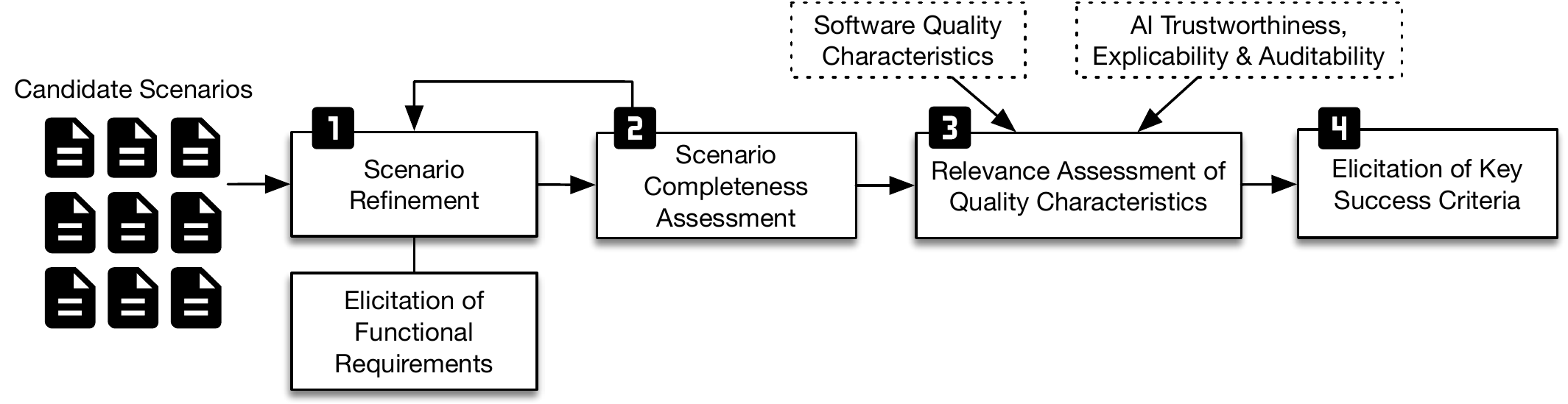}
\caption{Structure and process of the interview-based case study.} \label{fig:structure}
\end{figure}
Figure \ref{fig:structure} shows the structure and process of the case study. In total, we conducted 14 interviews with stakeholders from the 3 industry partners and from the 3 specialized SMEs for software development of AI-based systems. The numbers of interviewees per stakeholder role distributed as follows: DPO (2), SS (2), DS (3), MO (4), PLM (3). As in this paper our research concentrates on the individual relevance of quality characteristics and success criteria of the software platform, we only describe the results of steps 3 and 4 of the case study in more detail. To this end, we formulated the following three research questions:
\begin{itemize} 
\item \textbf{RQ1:} How do the stakeholders of the software platform assess the relevance of the ISO 25010:2011 (SQuaRE) \cite{25010:2011} characteristics for software quality, AI trustworthiness, explicability, and auditability?
\item \textbf{RQ2:} What are the key success factors of the stakeholders for human-AI teaming in smart manufacturing?
\item \textbf{RQ3:} What are potential metrics to evaluate these key success factors?
\end{itemize} 

Based on the guidelines by Runeson and H{\"o}st \cite{runeson_guidelines_2009}, we designed a questionnaire\footnote{https://bit.ly/3lV3aFw} for interviewing the stakeholders in step 3 and 4 of the case study regarding the relevance of quality characteristics and their success criteria towards the software platform. These interviews followed the refinement of the scenarios into functional requirements (step 1) and the completeness assessment of the scenarios (step 2). We also conducted a pilot interview as suggested by Yin \cite{yin_case_2017} with a highly experienced stakeholder and used his feedback to improve the questionnaire. Specifically, we refined definitions of quality characteristics in order to ensure a uniform level of understanding among the stakeholders. 

At the beginning of the interviews we explained the research context of our study - human-AI teaming in smart manufacturing - to the interviewees. Each interviewee had a thorough understanding of the research context since they have been participating in the project for over one year. Interviewees holding roles such as production line manager, data protection officer, and machine operator came from our industry partners. Likewise, interviewees holding roles such as software and data scientists came from the three specialized SMEs for software development of AI-based systems (cf. Section \ref{sec:researchcontext}). 

The questionnaire comprised two closed and one open questions. In the first closed question, we asked the interviewees to select the role (cf. Section \ref{roles}) that they most frequently perform. In order to not overlook any important stakeholder role, we deliberately asked them whether their most frequently performed role is on the presented list. The second closed question of the questionnaire examined the individual relevance of 11 quality characteristics, i.e., 8 quality characteristics of the ISO 25010:2011 (SQuaRE) \cite{25010:2011} standard for software quality and 3 AI-specific quality characteristics such as trustworthiness, explicability, and auditability. For easier reading, this question was divided into 11 sub-questions. To ensure common understanding of the quality characteristics, we presented the interviewees with a uniform definition of them. For the relevance assessment we adapted the \textit{Quality Attribute Workshop} format \cite{barbacci_quality_2003} and asked the interviewees to assign, in total, 100 points to the different quality characteristics according to their subjective relevance for human-AI teaming in smart manufacturing. In the final open question, we asked them to describe the key success factors in this context for their typical role. 

\begin{figure}[h!]
\includegraphics[width=\textwidth]{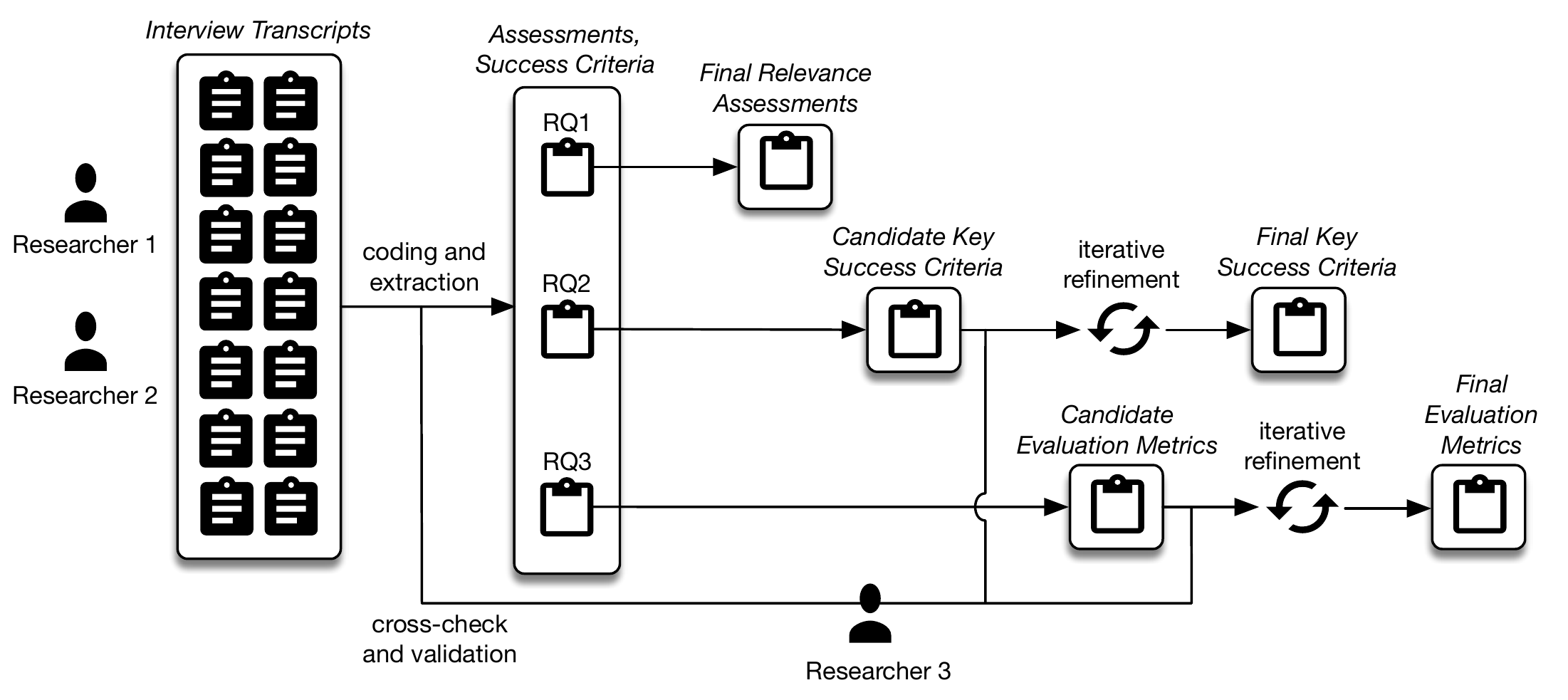}
\caption{Overview of research process and activities to answer the research questions.} \label{fig:researchProcess}
\end{figure}

Figure \ref{fig:researchProcess} shows the research process that structures the activities and specifies the outcome of each process step. The interviews were conducted by two researchers and the transcripts analyzed according to a predefined coding scheme. This scheme defined the coding and extraction of quantitative and qualitative data for each research question. The quantitative data related to RQ1 (relevance assessments of the quality characteristics) did not require further analysis. In the first step of analyzing the qualitative data from RQ2 and RQ3, two researchers highlighted the individual statements in the interview transcripts. After that, they iteratively refined the candidate key success criteria and evaluation metrics until they arrived at a consolidated set of criteria and metrics. A third researcher continuously checked and validated this refinement process. We repeated this process until we reached an agreement among all researchers.

\section{Results}
\label{sec:results}
In the following we present the results of our research questions, as defined in Section \ref{sec:researchquestions}.
\subsection{RQ1}
\begin{figure}[h!]
\begin{center}
\includegraphics[trim=68 0 65 2,clip,scale=0.8]{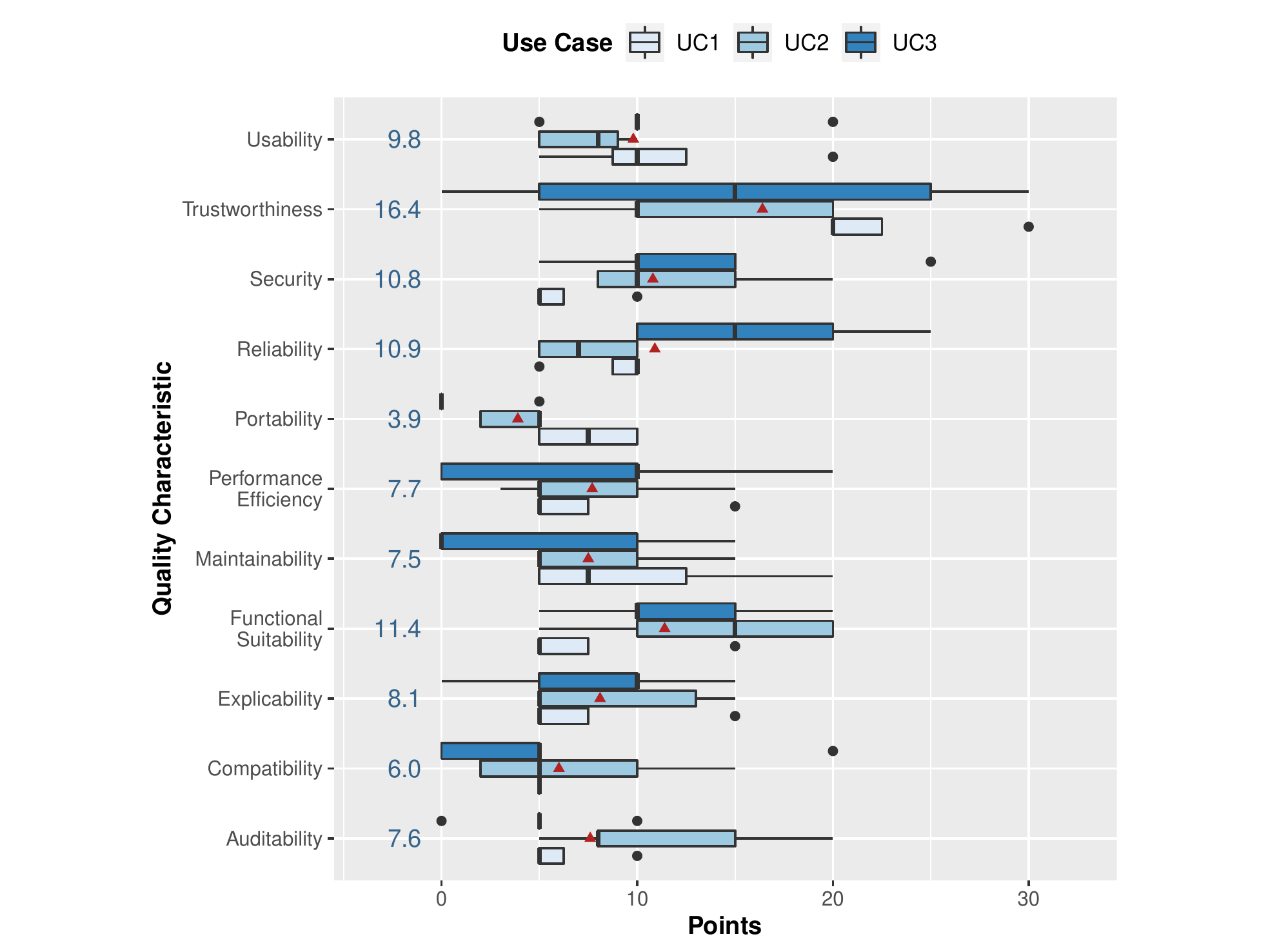}
\caption{Relevance assessments of the quality characteristics (per use case).} \label{fig:relevancePerUsecase}
\end{center}

\end{figure}

The results of this research question include the relevance ratings of the 11 quality characteristics by the interviewees. In Figure \ref{fig:relevancePerUsecase}, we show the rating results for each use case as well as the average rating for each quality characteristic. As shown in the illustration, the interviewees considered trustworthiness, functional suitability, and reliability as the most important quality characteristics for human-AI teaming. 

\begin{figure}[h!]
\begin{center}
\includegraphics[scale=0.88]{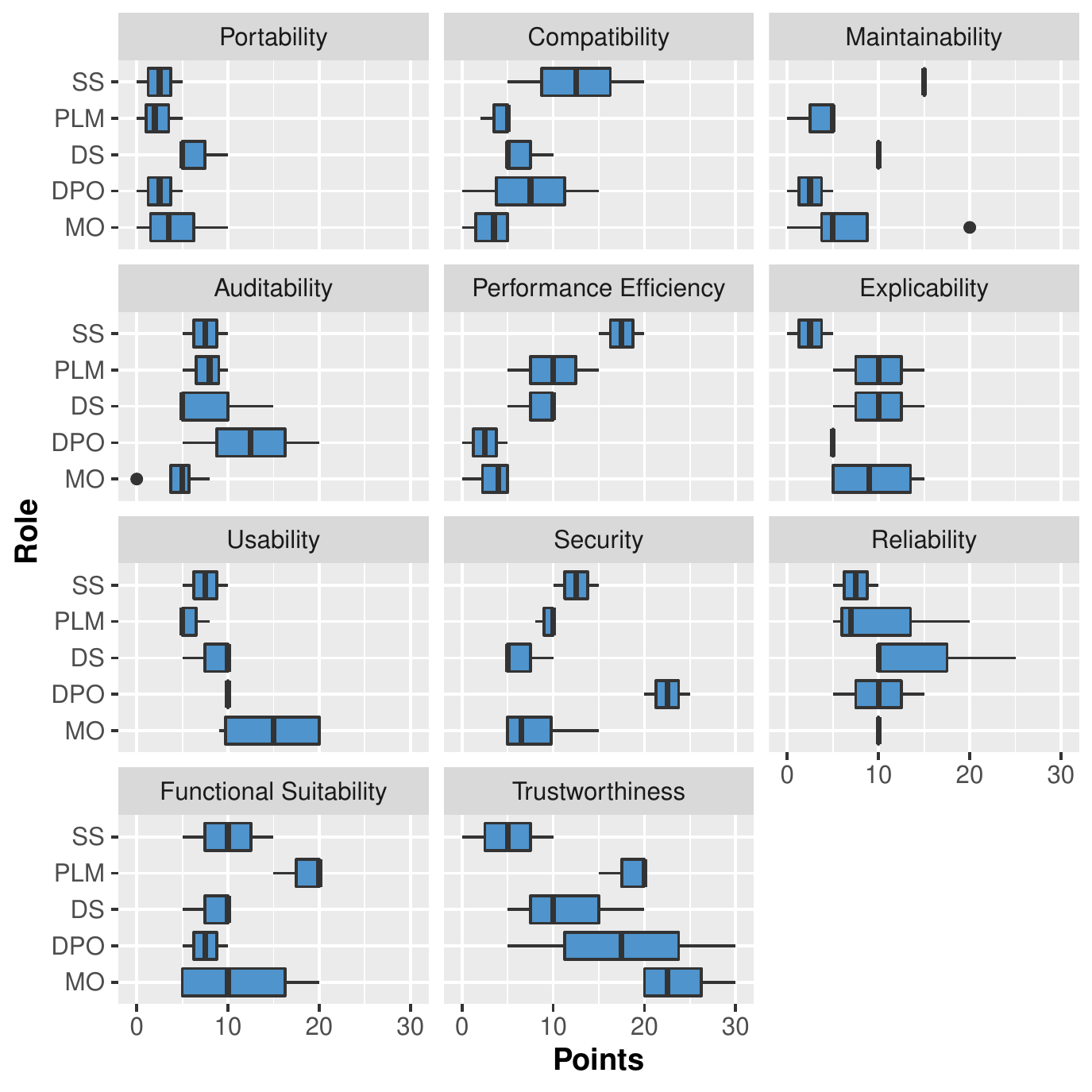}
\caption{Relevance assessments of the quality characteristics (per stakeholder role).} \label{fig:relevancePerStakeholder}
\end{center}
\end{figure}

Figure \ref{fig:relevancePerStakeholder} analyzes if the relevance assessments of the quality characteristics are also influenced by the stakeholder role (cf. Section \ref{roles}) of the interviewee. As we can see, each quality characteristic has a different relevance to each stakeholder role.
In the following, we present the two quality characteristics rated most relevant for each stakeholder role, with the average rating in brackets. \textbf{Software Scientist (SS)}: Performance Efficiency (17.5) and Maintainability (15); \textbf{Data Scientist: (SS):} Reliability (15) and Trustworthiness (11.7); \textbf{Data Protection Officer (DPO):} Trustworthiness (30) and Security (25); \textbf{Machine Operator (MO):} Trustworthiness (25) and Usability (10); \textbf{Production Line Manager (PLM):} Functional Suitability/Trustworthiness (ex aequo 18.3) and Reliability (10.7). In order to assess the dispersion of the relevance assessments, we finally calculated the standard deviation per quality characteristic: \textbf{Trustworthiness} (9.29), \textbf{Maintainability} (6.12), \textbf{Security} (6.04), \textbf{Functional Suitability} (6.02), \textbf{Performance Efficiency} (5.89), \textbf{Reliability} (5.71), \textbf{Compatibility} (5.57), \textbf{Auditability} (5.00), \textbf{Explicability} (4.86), \textbf{Usability} (4.84), \textbf{Portability} (3.39). In this context, the standard deviation can serve as a basic indicator of consensus or disagreement among the interviewees about the relevance of a quality characteristic. 
\subsection{RQ2 and RQ3}
Following, we summarize key success criteria for human-AI teaming in smart manufacturing and metrics for evaluating them for each stakeholder role. 

\subsubsection{Data Protection Officer (DPO):} The interviewees mentioned (a) Traceability of data processing, (b) Ensuring operator anonymity, and c) Ensuring operator and machine data confidentiality as key success criteria. We consider these identified success criteria to be functional requirements and did not formulate metrics for them. 

\subsubsection{Software Scientist (SS):} The condensed two key success criteria for this role cover a) Monitoring of realtime and historical production data, and b) Customizability of dashboards. Similary to the previous role, we regard these success criteria as functional requirements and abstained from formulating metrics.

\subsubsection{Data Scientist (DS):} In Table \ref{tab:ds} we enlist the condensed key success criteria from the perspective of data scientists. Similarly to the previous roles, not for all identified criteria meaningful metrics can be defined.
\subsubsection{Machine Operator:} For this role, we identified eight key success criteria and seven metrics for their evaluation. As shown in Table \ref{tab:mo}, the criteria and metrics tend to focus on time spans for failure detection and notification, as well as idle (waiting) times for either the machine or operator.
\begin{table}[h!]
\caption{Key success criteria and metrics for data scientists.\label{tab:ds}}
\begin{tabular}{P{0.7cm}p{5.6cm}p{5.6cm}}
\hline
\textbf{UC} & \textbf{Key success criteria} & \textbf{Metrics} \\ \hline
         1-3   &  Extensibility of data sources                           &         -        \\
         1-3   &  Customizability of dashboards                          &         -        \\ 
         1-3   &	Interoperability with Explainable \newline AI frameworks	 &  - \\
         3   &	Reliable production scene recognition & Scene recognition accuracy \\
         3   &	Reliable operator posture recognition	 & Operator posture recognition accuracy \\
            \hline
\end{tabular}
\end{table}

\begin{table}[h!]
\caption{Key success criteria and metrics for machine operators.\label{tab:mo}}
\begin{tabular}{P{0.7cm}p{5.6cm}p{5.6cm}}
\hline
\textbf{UC} & \textbf{Key success criteria} & \textbf{Metrics} \\ \hline
         1   &    Reduction of scrap rate                         &       Scrap rate          \\
         1   &     Shortening of production cycle time                        &   Production cycle time              \\ 
         1   &	Reliable prediction of faulty parts	 &  Faulty part prediction accuracy \\
         1   &	Facilitating root cause analysis	 &  - \\

         2   & Realtime detection of product \newline quality deviations  &   Time between product part analysis \newline and prediction result \\
         2   & Realtime notification of \newline production failures &    Time between detection and \newline notification of production failures             \\ 
         3 & Prevention of ergonomic risk & Freq. of unfavorable operator postures \\
         3 & Improvement of operator efficiency & Operator idle time \\
            \hline
\end{tabular}
\end{table}

\begin{table}[h!]
\caption{Key success criteria and metrics for production line managers.\label{tab:plm}}
\begin{tabular}{P{0.7cm}p{5.6cm}p{5.6cm}}
\hline
\textbf{UC} & \textbf{Key success criteria} & \textbf{Metrics} \\ \hline
         1   &         Reduction of scrap rate                    &      Scrap rate           \\
         1   &         Shortening of production cycle time                    &      Production cycle time           \\ 
         1   &		Reliable prediction of faulty parts & Faulty part prediction accuracy \\
         1-3   &         Improvement of OEE &   OEE               \\ 

         2   &         Realtime failure prediction                    &    Time between product part analysis \newline and failure notification             \\
         2   &         Shortening of machine downtimes                    &     Machine downtime            \\ 
         2   &		Shortening of machine idle times & Machine idle time \\

         3   &         Prevention of ergonomic risk                    &     Frequency of unfavorable operator \newline postures            \\
         3   &         Improvement of OLE                    &   OLE               \\ 
         3   &		Increasing operator satisfaction & Operator satisfaction score \\
            \hline
\end{tabular}
\end{table}

We use the frequency of particular unfavorable postures taken by the operator within the manufacturing process to assess the ergonomic risk. In this context, the notion of \textit{unfavorable postures} is taken from workplace safety methods (e.g., WISHA Caution Zone Checklist \cite{washington_state_department_of_labour_industries_hazard_2022}, RULA \cite{mcatamney_rula_1993}, REBA \cite{hignett_rapid_2000}, OWAS \cite{karhu_correcting_1977}). This includes, for example, awkward postures, heavy hand forces, repetitive motions, repeated impacts on the limbs, heavy or frequent lifting, and moderate to high hand-arm vibrations.
\subsubsection{Production Line Manager (PLM):} As depicted in Table \ref{tab:plm}, the majority of success criteria for this role focus on maximizing efficiency and effectiveness in the production process.

In this regard, \textit{Overall Equipment Effectiveness (OEE)} is used to determine how well machines are utilized in comparison to their potential. Similarly, \textit{Overall Labor Effectiveness (OLE)} quantifies the utilization, performance, and quality of the human workforce in the manufacturing process. In order to measure the satisfaction of the machine operator, we define the \textit{Operator Satisfaction Score} similarly to the \textit{System Usability Score (SUS)} \cite{brooke_sus_1996,brooke_sus_2013}. It measures operator satisfaction with specific aspects of the manufacturing process using Likert scales. This shall facilitate detecting changes in operator satisfaction as a result to changes in the production process.

\section{Discussion}
\label{sec:discussion}
From the perspective of the use cases, the quality characteristics related to the SQuaRE standard \cite{25010:2011}, portability, compatibility, and maintainability were rated as least relevant. Out of the three AI-specific quality characteristics, the interviewees rated auditability as the least important. Also it can be noted that the relevance of each quality characteristic is assessed differently for each use case. The use cases for parameter optimization (UC2) and large-scale parts assembly (UC3) directly affect the manufacturing process, whereas the use case for quality inspection (UC1) only supports the machine operator during quality inspection. The assessment results confirm this slightly different objective of the use cases, with functional suitability and security ranking less important for UC1 than for UC2 and UC3. However, we cannot identify a generic pattern that describes the connection between the use cases and their impact on the relevance assessments of each quality characteristic. 

The qualitative analysis of the interview responses revealed that some of the key success criteria are more closely related to functional requirements than to quality (non-functional) requirements. Broy  and Glinz \cite{glinz_rethinking_2005,glinz_non-functional_2007} already pointed out that there is often a lack of clarity in practice regarding the difference between functional and quality requirements. Unlike functional requirements, however, quality requirements can also be assessed by evaluating the extent to which they have been met. Therefore, we only defined metrics for key success criteria that are implicitly linked to quality requirements.

\section{Threats to Validity}
\label{sec:threats}
Different interpretations of the quality characteristics by the interviewees undermine the \textit{construct validity} of this study, which is primarily due to the fact that they have different roles and experiences. We tried to mitigate this threat by showing each interviewee a uniform definition of the quality characteristics that did not require any specialized knowledge. In addition, each interviewee was asked to raise any questions prior to the interview so that we could clarify any ambiguities. We also considered the role of each interviewee within the company when summarizing the interview answers, so we could determine from what perspective and with what intent each statement was delivered. 

Because our research project focuses on the applicability of AI in smart manufacturing, the greatest threat to \textit{internal validity} can be observed among interviewees, who tend to emphasize exclusively the AI-related quality characteristics. In our opinion, however, this threat is negligible, since as soon as we noticed this trend, we made the interviewee aware that overemphasizing one quality characteristic may result in underestimating the significance of others. In addition, only 100 points were available to distribute among the quality characteristics to reflect their relative importance.

As a final point, we recognize that the small sample size of interviewees in total might undermine the \textit{external validity} of our study. To mitigate this threat, we conducted interviews with different companies and with interviewees who hold different roles. Despite this, we see a threat to the generalizability of the results to other industries due to the functional and quality requirements their products must meet. 

\section{Conclusion and Future Work}
\label{sec:conclusion}
This paper presented the results of an interview-based case study to examine the relevance of 11 quality characteristics for human-AI teaming in smart manufacturing. The quality characteristics comprised the 11 characteristics of the ISO 25010:2011 standard for software quality (SQuaRE) and 3 AI-specific quality characteristics such as trustworthiness, explicability, and auditability. In the frame of an international research project, we develop an AI-based software platform that shall facilitate the cooperation between machine operators and manufacturing systems. For the presented case study, we conducted 14 interviews with stakeholders working in automotive industry, wind power plant assembly, and software development for AI-based software systems to assess the individual relevance of the 11 quality characteristics. Therefore, they were asked to distribute 100 points across the quality characteristics according to their relevance. 

The interviewees rated trustworthiness, functional suitability, reliability, and security as the most important quality characteristics, and portability, compatibility, and maintainability as the least important. Furthermore, the results indicate consensus regarding the relevance of the quality characteristics among interviewees with the same role. However, we also recognized that the relevance of the quality characteristics varies according to the concrete use case for the prospective software platform. Accordingly, we identified the improved production cycle efficiency, lower faulty parts and scrap, and reduced ergonomic risks as the key success criteria for human-AI teaming in smart manufacturing. The time span for detecting deviations (product or process quality), \textit{Overall Equipment Effectiveness (OEE)}, \textit{Overall Labor Effectiveness (OLE)}, the accuracy of fault prediction and scene recognition, and the accuracy of operator posture recognition are the most relevant metrics for evaluating these criteria.

Future research should focus on operationalizing these quality characteristics so that they can be continuously monitored during operation of AI-based smart manufacturing systems. In addition, an empirical study on the relevance of these quality characteristics is recommended after the interviewees have acquired experience with the prospective software platform. 
\section*{Acknowledgements}
This project has received funding from the European Union’s Horizon 2020 Research and Innovation Programme under grant agreement number 957402.

\bibliographystyle{splncs04}
\bibliography{bibliography.bib,standards.bib,method.bib,software.bib}

\end{document}